\documentclass[12pt,draft, onecolumn]{IEEEtran}

\usepackage{xspace,syntonly,color}
\usepackage{amsmath}
\usepackage{cite}
\usepackage{bm}
\usepackage{cases}
\usepackage{amssymb}
\usepackage{amsfonts}
\usepackage{graphicx}
\usepackage{graphics}
\usepackage{epsfig}
\usepackage{algorithm}
\usepackage{algorithmicx}
\usepackage{algpseudocode}
\usepackage{setspace}

\renewcommand{\baselinestretch}{2.0}

\psfull

\newtheorem{lemma}{\hspace{-11pt}\bf Lemma}
\newtheorem{theorem}{\hspace{-11pt}\bf Theorem}

\begin{document}

\title{\Large \sf \hspace{1\linewidth} Energy-Efficient Transmission Schedule for Delay-Limited Bursty Data Arrivals under Non-Ideal Circuit Power Consumption$^*$ \vspace{0.3cm}}

\author{
   {\normalsize \it \large Zheng Nan,}
   {\normalsize \it \large Tianyi Chen,}
   {\normalsize \it \large Xin Wang {{\small\sf (contact author)}},}
   {\normalsize \it \large and Wei Ni}\\
   \thanks{Work in this paper is supported by the China Recruitment Program of Global Young Experts, the Program for New Century Excellent Talents in University, and a research grant from Okawa Foundation.}
\thanks{Z. Nan, T. Chen and X. Wang are with the Dept. of Communication Science and Engineering, Fudan University, 220 Han Dan Road, Shanghai, China, email: ~xwang11{\rm\char64}fudan.edu.cn; W. Ni is with the CSIRO Computational Informatics (CCI), Sydney, Australia, NSW 2122.}
\thanks{Parts of the work have been presented at ICC 2014 \cite{ICC14} and ChinaSIP 2014 \cite{ChinaSIP14}.}
   }

\maketitle

\vspace*{-2.0cm}
\begin{abstract}
\noindent
This paper develops a novel approach to obtaining energy-efficient transmission schedules for delay-limited bursty data arrivals under non-ideal circuit power consumption. Assuming a-prior knowledge of packet arrivals, deadlines and channel realizations, we show that the problem can be formulated as a convex program. For both time-invariant and time-varying fading channels, it is revealed that the optimal transmission between any two consecutive channel or data state changing instants, termed epoch, can only take one of the three strategies: (i) no transmission, (ii) transmission with an energy-efficiency (EE) maximizing rate over part of the epoch, or (iii) transmission with a rate greater than the EE-maximizing rate over the whole epoch. Based on this specific structure, efficient algorithms are then developed to find the optimal policies that minimize the total energy consumption with a low computational complexity. The proposed approach can provide the optimal benchmarks for practical schemes designed for transmissions of delay-limited data arrivals, and can be employed to develop efficient online scheduling schemes which require only causal knowledge of data arrivals and deadline requirements.

\end{abstract}

\vspace*{-0.8cm}
   {\small
      \begin{center}
        \[
           \begin{array}{rl}
               \text{\bf Submission date:} & \text{\today}\\
           \end{array}
         \]
      \end{center}
    }

\markboth{}{}


\section{Introduction}

To prolong the operating lifetime of many battery powered commercial and tactical wireless (e.g., sensor) networks, energy-efficiency has appeared to be a critical issue. Energy-efficient resource allocation strategies were extensively pursued in \cite{Berry02, Uysal04, ElGamal02, Yao&Giannakis2005, Wan08IT}, where the goal is to minimize the transmission energy expenditure subject to average rate or delay constraints. Such an energy minimization is carried out over an infinite horizon and does not directly translate into quality-of-service (QoS) guarantees over finite time intervals. For QoS provisioning over finite time intervals, \cite{Uysal-Biyikoglu02net} considered minimizing the transmission energy for bursty packet arrivals with a single strict deadline. It was shown that a so-called lazy scheduling is the most energy-efficient by properly selecting minimum transmission rates for arriving packets under the causality constraints. Generalizing the lazy-scheduling, a calculus approach was proposed to find the optimal data departure curve (thus the optimal rate schedule) for packet arrivals with individual delay constraints, by the trajectory of letting a string tie its two ends and then taut between the data arrival and minimum departure curves \cite{Zafer&Modiano2005, Chen08}. The approaches in \cite{Uysal-Biyikoglu02net, Zafer&Modiano2005, Chen08} only apply to packet transmissions over time-invariant channels. Assuming a one-packet-per-slot arrival process and the same delay requirement for all packets, a recursive ``Constrained FlowRight'' algorithm was developed to find the energy-efficient scheduling over time-varying fading channels in \cite{Chen09}. For arbitrary packet arrival process and delay constraints, an efficient algorithm was put forth to find the optimal rate control strategy over time-varying wireless channels with a low computational complexity \cite{Wang13}.


All the works \cite{Berry02, Uysal04, ElGamal02, Yao&Giannakis2005, Wan08IT, Uysal-Biyikoglu02net, Zafer&Modiano2005, Chen08, Chen09, Wang13} assumed an ideal (negligible) circuit-power model. This holds for typical long-range transmissions. However, for short-range wireless (sensor) networks, non-ideal circuit power consumption due to signal processing (filters, DSP, oscillators, converters, etc.) needs to be taken into account; yet, there are few studies on the effects of the non-ideal circuit power on energy-efficient transmission policies for delay-limited data packets. In a different yet relevant context, \cite{Xu13, Bai11, Orh12} investigated sum-throughput maximization for packet transmissions over time-invariant channels subject to the causality and battery-capacity constraints due to an energy harvesting (arrival) process. However, these algorithms are inapplicable to addressing the critical issue of optimizing the energy efficiency for transmissions of delay-sensitive packets in general situations where energy harvesting does not take place and batteries are the only source of energy.

In this paper, we develop a novel unified approach to obtaining energy-efficient transmission schedules for bursty data packets with strict deadlines under the non-ideal circuit power consumption. Assuming that full knowledge of channel states, packet arrivals and deadlines is available a-prior, we consider the optimal (offline) policies that minimize the total energy consumption. Through a judicious convex formulation and the resultant optimality conditions, we reveal the structure of the optimal schedule. Specifically, we show that the optimal transmission between any two consecutive data or channel state changing instants (referred to as an epoch) can only take one of the three (``off'', ``on-off'', ``on'') strategies: (i) no transmission, (ii) transmission with the energy-efficiency (EE) maximizing rate $r_{ee}$ over a portion of the epoch, (iii) transmission with a rate $r>r_{ee}$ over the whole epoch. Based on this structure, we propose an efficient ``clipped string-tautening'' algorithm to find the optimal transmission policy with a low computational complexity for a time-invariant channel. Interestingly, it is shown that the calculus approach in \cite{Zafer&Modiano2005} can be modified to find the optimal policy; namely, the optimal data departure for the general non-ideal circuit-power case can be obtained by simply adjusting the ideal-case data departure in accordance to the EE-maximizing rate value. 
The proposed approach is then generalized to time-varying channels. In this case, it is shown that the optimal transmit-power allocation admits a multi-level water-filling form, where the water-levels can be obtained by a ``clipped water-tautening'' procedure. Our approach provides the optimal benchmarks for practical schemes designed for transmissions of delay-limited data arrivals over time-invariant and time-varying channels. It can be also employed to develop efficient online scheduling schemes which require only causal knowledge of channel states, data arrivals and deadline requirements.

The rest of the paper is organized as follows. Section~II describes the system models. Section~III and IV present the proposed approaches to energy-efficient transmissions of delay-limited bursty data packets over time-invariant and time-varying channels, respectively. Section~V provides the numerical results to evaluate the proposed schemes, followed by a conclusion in Section~VI.


\section{System Models}

Consider a wireless link with complex-valued baseband equivalent channel coefficient $h$. For simplicity, all nearby devices are supposed to use orthogonal channels so that interferences from other links are negligible. Assume without loss of generality (w.l.o.g.) that the noise at the receiver is a circularly symmetric complex Gaussian (CSCG) random variable with zero mean and unit variance. Given a transmit-rate $r$, we adopt the well-known Shannon-capacity formula as the minimum required transmit-power function:
\begin{equation}\label{eq2}
P(r)=\displaystyle \frac{1}{|h|^2}(e^r-1).
\end{equation}
Note that the Shannon formula is only used for specificity. It has been shown that with many modulation and coding schemes, transmit-power is an increasing and strictly convex function of the transmission rate. Our approach applies generally to any of these power functions $P(r)$.

\subsection{Data Arrival and Deadline Processes}

\begin{figure}[t]
\vspace{-0.4cm}
\centering
\includegraphics[width=4in]{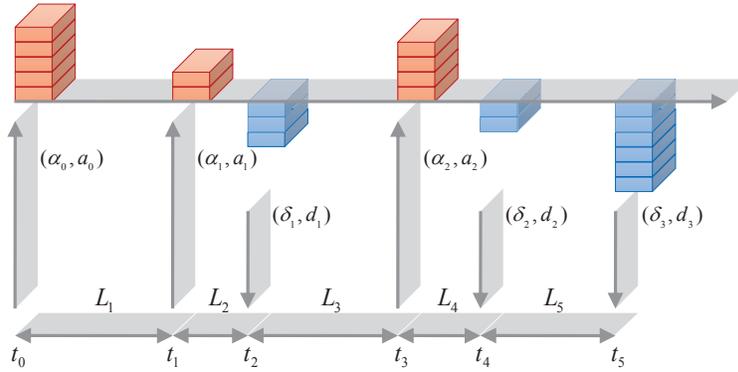}
\vspace{-0.8cm}
\caption{Data arrival and deadline processes.}\label{model}
\vspace{-0.5cm}
\end{figure}
Consider a wireless link with data packets transmitted from a transmitter to a receiver. We say that the data state changes when new data packets arrive or a data deadline is reached. As shown in Fig.~\ref{model}, over the entire transmission interval $[0, T]$, assume that there are $N+1$ data state changing instants $0=t_0<t_1<t_2<\cdots<t_N=T$. We refer to the time interval between two consecutive data state changing instants as an epoch; the length of the $n$th epoch is $L_n=t_n-t_{n-1}$, $n=1,\ldots,N$. 

The packet arrival process is modeled by a set $\mathcal{A}:=\{(\alpha_0, a_0), (\alpha_1, a_1),\ldots, (\alpha_A, a_A)\}$, as illustrated by the red bricks in Fig.~1, where $A$ denotes the number of data arrival events, $\alpha_i$ denotes the epoch index of the $i$th arrival time, and $a_i$ denotes the number of packets arriving at $t_{\alpha_i}$, $i=0,\ldots,|A|$, where $|\cdot|$ denotes cardinality. Let $\mathbb{N}$ and $\mathbb{N}^+$ denote the sets of non-negative and positive integers. Clearly, we have $\alpha_i \in \mathbb{N}$. Let $\boldsymbol{\alpha}:=\{\alpha_0, \alpha_1, \ldots, \alpha_A\}$ with $0=\alpha_0 < \alpha_1 < \cdots < \alpha_A = N$ for convenience. Let $\boldsymbol{a}:=\{a_0, a_1, \ldots, a_A\}$. For the number of arrivals at $\alpha_i$, we clearly have $a_i \in \mathbb{N}^+$. Since the last time instant for packet transmission is $N-1$, no packets should be allowed to arrive at $\alpha_A = N$; hence, we must have $a_A =0$. We include the pair $(\alpha_A, a_A) = (N, 0)$ for the ease of problem formulation and algorithm development.

The deadline requirements for the packets are described by another set $\mathcal{D}:=\{(\delta_1, d_1), (\delta_2, d_2),\ldots, \linebreak (\delta_D, d_D)\}$, as illustrated by the blue bricks in Fig.~1, where $D$ denotes the number of deadlines, $\delta_j$ denotes the epoch index of the $j$th deadline, and $d_j$ denotes the number of packets which should depart within $(t_{\delta_{j-1}},t_{\delta_j}]$ (we let $t_{\delta_0}=0$ for convenience). Let $\boldsymbol{\delta}:=\{\delta_1, \ldots, \delta_D\}$, with $0< \delta_1 < \cdots < \delta_D = N$ and $\delta_j \in \mathbb{N}^+$. Let $\boldsymbol{d}:=\{d_1, \ldots, d_D\}$ where $d_j \in \mathbb{N}^+$. The total number of data packets arriving and transmitted over time interval $[0,T]$ is obviously $G:=\sum_{i=0}^{A-1}a_i=\sum_{j=1}^{D}d_j$.

\subsection{Non-Ideal Circuit Power Consumption}

In short-range wireless networks, circuit power consumption for e.g. the AC/DC converter and radio frequency (RF) amplifier is non-negligible when transmit-power $P>0$. When there is no data transmission, the transmitter could turn off the power amplifier and switch into a micro-sleep mode to avoid/reduce the circuit power consumption \cite{Blu10}. For the ease of description, we refer to the transmitter status with a transmit-power $P >0$ and that with $P = 0$ as the ``on'' and ``off'' modes, respectively. Let $\rho \geq 0$ denote the circuit power during the ``on'' mode, $\eta \in (0, 1]$ denote the efficiency of the RF chain, and $\beta \geq 0$ the circuit power consumed during the ``off'' mode. The total power $P_{total}$ consumed by a transmitter is then \cite{Xu13, Mia10}:
\begin{equation}\label{eq1}
P_{total}=\left\{
\begin{array}{ll}
    \frac{P}{\eta}+\rho,		&P>0,\\
    \beta,			&P=0.\\
\end{array}
\right.
\end{equation}
In practical systems, $\beta$ is usually much smaller compared to $\rho$ \cite{Mia10} and thus can be neglected for simplicity. Hence, we can assume w.l.o.g. the circuit-power during the ``on'' and ``off'' modes to be $\rho>0$ Watts and $\beta=0$ Watt, respectively. We further assume $\eta=1$ w.l.o.g since $\eta$ is only a scaling constant.


\section{Time-Invariant Channel}

Consider first a static channel with time-invariant channel coefficient $h$. Due to the non-ideal circuit power consumption, the transmission can be turned on for only a portion of an epoch and turned off afterwards to save energy. Let $\boldsymbol{l}^{on}=\{l_1^{on}, l_2^{on}, \ldots, l_N^{on}\}$ collect the ``on'' periods with length $0 \leq l_n^{on} \leq L_n$ in the $n$th epoch. Given that the power function $P(r)$ is convex, it was proved that the transmit-rate over the ``on'' period $l_n^{on}$ of each epoch $n$ should remain unchanged in the optimal policy \cite{Xu13}. Let $\boldsymbol{r}:=\{r_1, r_2, \ldots, r_N\}$ collect such invariant transmit-rates over the ``on'' period of each epoch.
For a bursty data arrival process modeled by ($\mathcal{A}$, $\mathcal{D}$), the energy-efficient transmission schedule is to select an optimal set of $\{\boldsymbol{r},\;\boldsymbol{l}^{on}\}$ such that the total energy consumed for delivery of the arriving data packets ahead of deadlines is minimized; i.e., we wish to solve:
\footnote{The problem formulation of the more general cases with $\beta>0$ and $\eta<1$ can be transformed into a similar form, where the objective $\min_{\boldsymbol{r}, \boldsymbol{l}^{on}}{\sum_{n=1}^N{[(\frac{P(r_n)}{\eta}+\rho) l_n^{on}+\beta(L_n-l_n^{on})]}}$ is equivalent to $\min_{\boldsymbol{r}, \boldsymbol{l}^{on}}{\sum_{n=1}^N{[(P(r_n)+\eta (\rho-\beta)) l_n^{on}]}}$. Hence, our results readily carry over to such cases by simply involving $\rho \equiv \eta (\rho-\beta)$.}
\begin{equation}\label{eq3}
\begin{array}{cll}
\displaystyle\min_{\boldsymbol{r},\;\boldsymbol{l}^{on}} &\displaystyle \sum_{n=1}^{N}{[(P(r_n)+\rho)l_{n}^{on}]}\\
\text{s.t.} &\text{(C1):} \;\; \displaystyle\sum_{n=1}^{\alpha_i}{(r_n l_n^{on}) \leq \sum_{k=0}^{i-1}{a_k}}, & i=1, \ldots, A,\\
&\text{(C2):} \;\; \displaystyle\sum_{n=1}^{\delta_j}{(r_n l_n^{on}) \geq \sum_{k=1}^{j}{d_k}}, & j=1, \ldots, D,\\
&{\text{(C3):} \;\; \displaystyle r_n \geq 0, \;\; 0 \leq l_n^{on} \leq L_n}, & n=1, \ldots, N.
\end{array}
\end{equation}
Here, in addition to the trivial constraints (C3), (C1) presents the causality constraints: the number of packets $\sum_{n=1}^{\alpha_i}{(r_nl_n^{on})}$ transmitted before the $i$th arrival time instant must not exceed the number of available packets $\sum_{k=0}^{i-1}{a_k}$ in the transmit buffer. (C2) presents the deadline constraints: the number of packets $\sum_{n=1}^{\delta_j}{(r_nl_n^{on})}$ transmitted before the $j$th deadline should be no less than the required number of packets $\sum_{k=1}^{j}{d_k}$. 


\subsection{Convex Reformulation and Optimality Conditions}

In the ideal circuit-power ($\rho=\beta=0$) case, it was shown that the transmitter is always on (i.e., $l_n^{on*}=L_n$) in the optimal policy \cite{Berry02, Uysal04, ElGamal02, Yao&Giannakis2005, Wan08IT, Uysal-Biyikoglu02net, Zafer&Modiano2005, Chen08, Chen09, Wang13, Xu13}. The optimal transmission schedule then reduces to an optimal rate control problem. With $\boldsymbol{r}$ as the only optimization variable, \eqref{eq3} is a convex program as long as $P(r_n)$ is convex. However, in the general non-ideal circuit-power case, $\boldsymbol{l}^{on}$ is also a variable to be optimized. Since both $P(r_n)l_{n}^{on}$ and $r_n l_n^{on}$ are neither concave nor convex in $(r_n,  l_n^{on})$, the problem \eqref{eq3} is non-convex. Yet, we next show that it can be reformulated into a convex program through a change of variables.

Define $\Phi_n:=r_n l_n^{on}$. With $\boldsymbol{\Phi}:=\{\Phi_1, \ldots, \Phi_N\}$, we rewrite \eqref{eq3} as:
\begin{equation}\label{eq4}
\begin{array}{cll}
\displaystyle\min_{\bm{\Phi},\bm{l}^{on}} &\displaystyle\sum_{n=1}^{N}{[(P(\displaystyle \frac{\Phi_n}{l_n^{on}})+\rho)l_{n}^{on}]}\\
\text{s.t.} &\displaystyle\sum_{n=1}^{\alpha_i}{\Phi_n} \leq \sum_{k=0}^{i-1}{a_k}, &i=1, \ldots, A,\\
&\displaystyle\sum_{n=1}^{\delta_j}{\Phi_n} \geq \sum_{k=1}^{j}{d_k}, &j=1, \ldots, D,\\
&\displaystyle \Phi_n \geq 0, \quad 0 \leq l_n^{on} \leq L_n, &n=1, \ldots, N,
\end{array}
\end{equation}
where we define $P(\frac{\Phi_n}{l_n^{on}})l_n^{on}=0$ if $l_n^{on}=0$. For any convex $P(r_n)$, $P(\frac{\Phi_n}{l_n^{on}})l_n^{on}$ is called its perspective, which is a jointly convex function of $(\Phi_n, l_n^{on})$ \cite{convex, Wang11}. Since the constraints are all linear, it then readily follows that \eqref{eq4} is a convex problem.

Let $\boldsymbol{\Lambda}:=\{\lambda_i,i=1,\ldots,A,\;\mu_j,j=1,\ldots,D\}$ where $\lambda_i$ and $\mu_j$ denote the Lagrange multipliers associated with the causality and deadline constraints, respectively. The Lagrangian of \eqref{eq4} is given by:
\begin{equation}\label{eq5}
\begin{aligned}
\mathcal{L}(\boldsymbol{r},&\boldsymbol{l}^{on},\boldsymbol{\Lambda})=\displaystyle \sum_{n=1}^{N}{[(P(\displaystyle \frac{\Phi_n}{l_n^{on}})+\rho)l_n^{on}]}    +\sum_{i=1}^{A}{\lambda_i(\sum_{n=1}^{\alpha_i}{\Phi_n- \sum_{k=0}^{i-1}{a_k}})}    -\sum_{j=1}^{D}{\mu_j(\sum_{n=1}^{\delta_j}{\Phi_n}-\sum_{k=1}^{j}{d_k})}\\
\displaystyle = & \mathcal{C}(\boldsymbol{\Lambda})+\sum_{n=1}^{N}{[(P(\frac{\Phi_n}{l_n^{on}})+\rho)l_n^{on}-(\sum_{j=j_n}^{D}{\mu_j}-\sum_{i=i_n}^{A}{\lambda_i})\Phi_n]}\\
\end{aligned}
\end{equation}
where we define $i_n:=\arg \min{\{i:n \leq \alpha_i\}}$, $j_n:=\arg \min{\{j:n \leq \delta_j\}}$, and $\mathcal{C}(\boldsymbol{\Lambda}):=-\sum_{i=1}^{A}{\lambda_i}$ $(\sum_{k=0}^{i-1}{a_k})+\sum_{j=1}^{D}{\mu_j(\sum_{k=1}^{j}{d_k})}$.

Let $(\boldsymbol{\Phi}^*,\boldsymbol{l}^{on*})$ denote the optimal solution for \eqref{eq4} and $\boldsymbol{\Lambda}^*$ the optimal Lagrange multiplier vector for its dual problem. Upon defining $w_n:=\sum_{j=j_n}^{D}{\mu_j^*}-\sum_{i=i_n}^{A}{\lambda_i^*}$, we can derive from the Karush-Kuhn-Tucker (KKT) optimality conditions that: $\forall{n}$,
\begin{equation}\label{eq6}
\begin{aligned}
\displaystyle (\Phi_n^*, l_n^{on*})=\arg \min \;&{[(P(\displaystyle \frac{\Phi_n}{l_n^{on}})+\rho)l_n^{on}-w_n\Phi_n]}\\
\text{s.t.} \quad &\Phi_n \geq 0, \quad 0 \leq l_n^{on} \leq L_n.\\
\end{aligned}
\end{equation}
In addition, the non-negative Lagrange multipliers $\lambda_i^*$ and $\mu_j^*$ satisfy the complementary slackness conditions:
\begin{equation}\label{eq7}
\left \{
\begin{array}{c}
\lambda_i^*=0, \text{ if } \sum_{n=1}^{\alpha_i}{\Phi_n^*} < \sum_{k=0}^{i-1}{a_k},\\
\sum_{n=1}^{\alpha_i}{\Phi_n^*}=\sum_{k=0}^{i-1}{a_k}, \text{ if } \lambda_i^* > 0;
\end{array}
\right.
\;\;i=1, \ldots, A.
\end{equation}
\begin{equation}\label{eq8}
\left \{
\begin{array}{c}
\mu_j^*=0, \text{ if } \sum_{n=1}^{\delta_j}{\Phi_n^*} > \sum_{k=1}^{j}{d_k},\\
\sum_{n=1}^{\delta_j}{\Phi_n^*} = \sum_{k=1}^{j}{d_k}, \text{ if } \mu_j^* > 0;
\end{array}
\right.
\;\;j=1, \ldots, D.
\end{equation}

Let $r_n^*=\frac{\Phi_n^*}{l_n^{on*}}$ if $l_n^{on*}>0$, and $r_n^*$ take an arbitrary non-negative value when $l_n^{on*}=0$, $\forall{n}$. It is obvious that $(\boldsymbol{r}^*,\boldsymbol{l}^{on*})$ is the optimal solution to \eqref{eq3}. From \eqref{eq6}--\eqref{eq8}, the sufficient and necessary optimality conditions for \eqref{eq3} are: 
\begin{equation}\label{eq9}
\left.
\begin{split}
\displaystyle (r_n^*, l_n^{on*})=\arg \min \; & {[P(r_n)+\rho-w_n r_n]l_n^{on}}\\
\text{s.t.} \quad & r_n \geq 0, \;\; 0 \leq l_n^{on} \leq L_n;\\
\end{split}
\right.
\;\; \forall n.
\end{equation}
\begin{equation}\label{eq10}
\left \{
\begin{array}{c}
\lambda_i^*=0, \text{ if }\sum_{n=1}^{\alpha_i}{(r_n^*l_n^{on*})} < \sum_{k=0}^{i-1}{a_k},\\
\sum_{n=1}^{\alpha_i}{(r_n^* l_n^{on*})}=\sum_{k=0}^{i-1}{a_k}, \text{ if }\lambda_i^* > 0;
\end{array}
\right.
\;\;i=1, \ldots, A.
\end{equation}
\begin{equation}\label{eq11}
\left \{
\begin{array}{c}
\mu_j^*=0, \text{ if } \sum_{n=1}^{\delta_j}{(r_n^*l_n^{on*})} > \sum_{k=1}^{j}{d_k},\\
\sum_{n=1}^{\delta_j}{(r_n^*l_n^{on*})} = \sum_{k=1}^{j}{d_k}, \text{ if } \mu_j^* > 0;
\end{array}
\right.
\;\;j=1, \ldots, D.
\end{equation}

\subsection{Optimal Transmission Schedule}

Next, we develop an efficient algorithm to find the optimal $(\boldsymbol{r}^*,\boldsymbol{l}^{on*})$ satisfying \eqref{eq9}--\eqref{eq11}. Let $P'(r_n)$ denote the first derivative of $P(r_n)$. For any $l_n^{on}>0$, we can derive from \eqref{eq9} that
\begin{equation}\label{eq12}
\displaystyle r_n^*=\arg \min_{r_n \geq 0}{[P(r_n)+\rho-w_n r_n]}.
\end{equation}
As $P(r_n)$ is strictly convex and increasing, this is equivalent to: $P'(r_n^*)=w_n$. Substituting it into \eqref{eq9} implies:
\begin{equation}\label{eq13}
\displaystyle l_n^{on*}=\displaystyle \arg \min_{0 \leq l_n^{on} \leq L_n}{l_n^{on}[P(r_n^*)+\rho-P'(r_n^*)r_n^*]}.
\end{equation}

Now we consider a bits-per-Joule EE-maximizing rate:
\begin{equation}\label{eq14}
\displaystyle r_{ee}=\arg \max_{r \geq 0}{\displaystyle \frac{r}{P(r)+\rho}}=\arg \min_{r \geq 0}{\displaystyle \frac{P(r)+\rho}{r}}.
\end{equation}
Note that since $\frac{P(r)+\rho}{r}$ is a (convex-over-linear) quasi-convex function, it has a unique minimizer $r_{ee}$, which can be efficiently obtained by a simple bisectional search \cite{Xu13}.

Interestingly, we can rely on \eqref{eq13} to show that the optimal schedule depends on the EE-maximizing rate $r_{ee}$:
\begin{lemma}
{\it The optimal transmission policy for (\ref{eq3}) can only adopt one of the following three (``off'', ``on-off'' and ``on'') strategies per epoch $n$: (i) $l_n^{on*}=0$, (ii) $r_n^{on*}=r_{ee}$, $l_n^{on*} \leq L_n$, or (iii) $r_n^*>r_{ee}$, $l_n^{on*}=L_n$.}
\end{lemma}

\begin{IEEEproof}
See Appendix~A.
\end{IEEEproof}

Lemma 1 dictates that any transmit-rate $r_n<r_{ee}$ should not be adopted in the optimal policy. In fact, since $r_{ee}$ maximizes the bits-per-Joule EE, we can show that a transmission strategy with an $r_n<r_{ee}$ over an epoch is always dominated by an on-off transmission with $r_{ee}$, which can use less energy to deliver the same data amount. Only when the data deadlines are strict (i.e., no further delay is allowed) should we adopt an $r_n^* > r_{ee}$; in this case, the transmitter should be always on, i.e., $l_n^* = L_n$, over epoch $n$.

Let $P'^{-1}$ denote the inverse function of $P'$. We can obtain from \eqref{eq12} that
\begin{equation}
r_n^*=  \arg \min_{r_n \geq 0}{[P(r_n)+\rho-w_n r_n]}:=P'^{-1}(w_n) =  \log(|h|^2 w_n) 
\end{equation}
%
%
which is an increasing function of $w_n$. Using this fact and the complementary slackness conditions \eqref{eq10}--\eqref{eq11}, we can then establish that:
\begin{lemma}
{\it In the optimal policy, the rate $r_n^*$ can only change at $t_{\alpha_i}$ or $t_{\delta_j}$ where the causality or deadline constraints are met with equality; specifically, the rate increases after a $t_{\alpha_i}$ where $\sum_{n=1}^{\alpha_i}{(r_n^* l_n^{on*})}=\sum_{k=0}^{i-1}{a_k}$, and it decreases after a $t_{\delta_j}$ where $\sum_{n=1}^{\delta_j}{(r_n^* l_n^{on*})}=\sum_{k=1}^{j}{d_k}$.}
\end{lemma}
\begin{IEEEproof}
See Appendix~B.
\end{IEEEproof}

Lemma 2 reveals that the optimal rate control policy follows a specific pattern. Due to the convexity of rate function $P(r)$, a constant transmit-rate should be maintained whenever possible, to minimize the total energy consumption. In the optimal policy, the rate needs to be changed only when the data causality or deadline constraints become active. A causality constraint is active, i.e., all available data is cleared up at $t_{\alpha_i}$ when the amount of data arrivals so far is small; as a result, a lower rate is maintained before $t_{\alpha_i}$ than after. Similarly, a deadline constraint is active at $t_{\delta_j}$ when the deadline requirements are strict, thus a higher rate is maintained before $t_{\delta_j}$ than after. This is in the same spirit with the ``string tautening'' calculus approach developed in \cite{Zafer&Modiano2005}.

Based on the rules revealed in Lemmas~1--2, we then put forth an $r_{ee}$-clipped ``string tautening'' procedure in Algorithm~1 to construct the optimal policy.

%
%
%
%
%
%
\begin{spacing}{1.13}
\small
\vspace{0.25 in}
\hrule height 0.1pt depth 0.3pt
\vspace{0.05 in}
\noindent {\bf Algorithm~1} $r_{ee}$-Clipped ``String Tautening''
\vspace{0.05 in}
\hrule height 0.1pt depth 0.3pt
\vspace{0.1 in}
\begin{algorithmic}[1]
\Procedure{ScheduleR}{$\mathcal{A},\mathcal{D}$}
\State $N_{\text{offset}}=0$, and $r_n^*=0$, $\forall{n}$;
\While{$N_{\text{offset}}<N$}
	\State [$\tau$, $r$, $\Delta$]=FirstChangeR($\mathcal{A},\mathcal{D}$);
    \State find a set of $l_n^{on*}$ satisfying $\sum_{n=1}^{\tau}{(r l_n^{on*})}=\Delta$;
	\For{$n=1$ to $\tau$}
		\State $r_{N_{\text{offset}}+n}^*=r$, $l_{N_{\text{offset}}+n}^{on*}=l_n^{on*}$;
	\EndFor
	\State $N_{\text{offset}} = N_{\text{offset}} +\tau$;
	\State update $(\mathcal{A},\mathcal{D})$;

%
\EndWhile
\EndProcedure
\Statex
\Function{[$\tau$, $r$, $\Delta$]=FirstChangeR}{$\mathcal{A},\mathcal{D}$}
	\State sort $\alpha_i, \forall i$, and $\delta_j, \forall j$, together in ascending order into a vector $\pi:=\{\pi_1, \ldots, \pi_{A+D}\}$;
	\State $r^{-}=0$, $r^{+}=\infty$, $\tau^{-}=\tau^{+}=0$, $\Delta^- =\Delta^+ =0$;
	\For{$m=1$ to $A+D$}
		\If{$\pi_m=\alpha_i \in \boldsymbol{\alpha}$ for a certain $i$}
			\State $r_{\alpha_i}^+=\max{\{r_{ee}, \frac{\sum_{k=0}^{i-1}{a_k}}{\sum_{n=1}^{\alpha_i}{L_n}}\}}$;
			\If{$r_{\alpha_i}^+ \leq r^{+}$}
				\State $\tau^{+}=\alpha_i$, $r^{+}=r_{\alpha_i}^+$, $\Delta^+=\sum_{k=0}^{i-1}{a_k}$;
			\EndIf
		\EndIf

		\If{$\pi_m=\delta_j \in \boldsymbol{\delta}$ for a certain $j$}
			\State $r_{\delta_j}^-=\max{\{r_{ee}, \frac{\sum_{k=1}^{j}{d_k}}{\sum_{n=1}^{\delta_j}{L_n}}\}}$;
			\If{$r_{\delta_j}^- \geq r^{-}$}
				\State $\tau^{-}=\delta_j$, $r^{-}=r_{\delta_j}^-$, $\Delta^-=\sum_{k=1}^{j}{d_k}$;
			\EndIf
		\EndIf

		\If{$r^{-} > r^{+}$ \& $\tau^{-}<\tau^{+}$}
			\State return $\tau=\tau^{-}$, $r=r^{-}$, $\Delta = \Delta^-$;
		\ElsIf{$r^{-} \geq r^{+}$ \& $\tau^{-} \geq \tau^{+}$}
			\State return $\tau=\tau^{+}$, $r=r^{+}$, $\Delta = \Delta^+$;
		\EndIf
	\EndFor
\EndFunction
\end{algorithmic}
\vspace{0.05 in}
\hrule height 0.1pt depth 0.3pt
\vspace{0.25 in}
\end{spacing}
\vspace{0.2 in}

The key component in Algorithm~1 is the function FirstChangeR, which relies on Lemmas~1--2 to determine the first rate-changing time $t_{\tau}$ and the invariant rate $r$ used before it in the optimal policy for the $(\mathcal{A},\mathcal{D})$ system. In this function, $\tau^{+}$ and $\tau^{-}$ denote the epoch indices for the two candidate first rate-changing time instants, whereas $r^{+}$ and $r^{-}$ denote the candidate rates that are maintained over $[0,t_{\tau^{+}}]$ or $[0,t_{\tau^{-}}]$.

Suppose that a constant transmit-rate $r_n^*=r_{\alpha_i}^+$, $\forall{n \leq \alpha_i}$, is maintained in the optimal policy such that the corresponding $i$th causality constraint is met with equality at $t_{\alpha_i}$, i.e., $\sum_{n=1}^{\alpha_i}{(r_n^* l_n^{on*})}=\sum_{k=0}^{i-1}{a_k}$. By Lemma~1, $r_{\alpha_i}^+ \geq r_{ee}$ holds, and an $r_{\alpha_i}^+>r_{ee}$ renders $l_n^{on*}=L_n$, $\forall{n \leq \alpha_i}$. This implies that the packets $\sum_{k=0}^{i-1}{a_k}$ can only be delivered at $t_{\alpha_i}$ by either (i) a transmission with $r_{ee}$ over the ``on'' periods of a total length $\sum_{n=1}^{\alpha_i}{l_n^{on*}}=\frac{\sum_{k=0}^{i-1}{a_k}}{r_{ee}} \leq \sum_{n=1}^{\alpha_i}{L_n}$, or (ii) a transmission with a rate $\frac{\sum_{k=0}^{i-1}{a_k}}{\sum_{n=1}^{\alpha_i}{L_n}}>r_{ee}$ over the entire interval $[0, \alpha_i]$ of length $\sum_{n=1}^{\alpha_i}{L_n}$, if $\sum_{k=0}^{i-1}{a_k}>r_{ee}\sum_{n=1}^{\alpha_i}{L_n}$. In a simpler form, we have $r_{\alpha_i}^+=\max{\{r_{ee}, \frac{\sum_{k=0}^{i-1}{a_k}}{\sum_{n=1}^{\alpha_i}{L_n}}\}}$. Similarly, if a constant transmit-rate $r_n^*=r_{\delta_j}^-$, $\forall{n \leq j}$, is maintained such that the $j$th deadline constraint is met with equality at $t_{\delta_j}$, we must have $r_{\delta_j}^-=\max{\{r_{ee}, \frac{\sum_{k=1}^{j}{d_k}}{\sum_{n=1}^{\delta_j}{L_n}}\}}$.

In the function FirstChangeR, $r^{+}$ is updated as $r^{+}=\min{\{r^{+},r_{\alpha_i}^+\}}$, yielding $r^{+}=\min{\{r_{\alpha_1}^{+}, \ldots, r_{\alpha_i}^+\}}$. Note that $r_{\alpha_i}^+$ is in fact the upper bound for an invariant rate that can be used to satisfy the $i$th causality constraint. Hence, $r^{+}=\min_{\alpha_k \leq \alpha_i}{r_{\alpha_k}^{+}}$ is the maximum value for an invariant rate to satisfy all the causality constraints so far. Similarly, $r^{-}=\max_{\delta_k \leq \delta_j}{r_{\delta_k}^{-}}$ is the minimum rate to satisfy all the deadline constraints so far. At a certain $t_{\alpha_i}$ or $t_{\delta_j}$, if we have $r^{+}<r^{-}$, then there does not exist an invariant rate to satisfy all the causality and deadline constraints, i.e., the rate needs to be changed before this specific $t_{\alpha_i}$ or $t_{\delta_j}$. The first rate-changing time instant is obtained by simply comparing $\tau^{+}$ with $\tau^{-}$ to find which type of constraint first becomes active. 

If the returned $t_{\tau}<T$, we reuse Function FirstChangeR for a new $(\mathcal{A}, \mathcal{D})$ system over the remaining time to find the next rate-changing time and the next optimal transmit-rate. The update of the new $(\mathcal{A}, \mathcal{D})$ needs to take into account the time offset as well as the adjustment of $\boldsymbol{a}$ and $\boldsymbol{d}$ based on the data amount that has been delivered. All the rate-changing time instants and the corresponding transmit-rates can be determined by repeatedly calling Function FirstChangeR, until the entire optimal policy is obtained.

The global optimality and efficiency of the proposed Algorithm 1 are formally stated in the following theorem:

\begin{theorem}
{\it Algorithm 1 computes the optimal transmission policy for (\ref{eq3}) with a linear complexity ${\cal O}(A+D)$.}
\end{theorem}

\begin{IEEEproof}
See Appendix~C.
\end{IEEEproof}

We prove Theorem 1 by showing the existence of a Lagrange multiplier vector $\boldsymbol{\Lambda}^*$, with which $\boldsymbol{r}^*$ and $\boldsymbol{l}^{on*}$ satisfy the sufficient and necessary optimality conditions \eqref{eq9}--\eqref{eq11}. The global optimality of $\{\boldsymbol{r}^*,\boldsymbol{l}^{on*}\}$  thus follows. In the search of the rate-changing points in Algorithm 1, we only need to go through the $A+D$ data arrival or deadline time instants, leading to a complexity ${\cal O}(A+D)$. 

Relying on the optimality conditions to directly construct the optimal solution for the problem at hand, the proposed Algorithm~1 is much more efficient than general solvers such as the (iterative) interior point methods\footnote{The interior point methods typically have a complexity higher than ${\cal O}(N^3)$ per iteration.} in terms of computational complexity. This is also corroborated by our simulations, which indicate that the CPU time for Algorithm~1 to obtain the optimal schedule can be less than 0.01\% of that with the standard CVX program \cite{cvx}.

\subsection{Visualization of the Optimal Policy}

\begin{figure}[t]
\vspace{-0.3cm}
\centering
\includegraphics[width=4.2in]{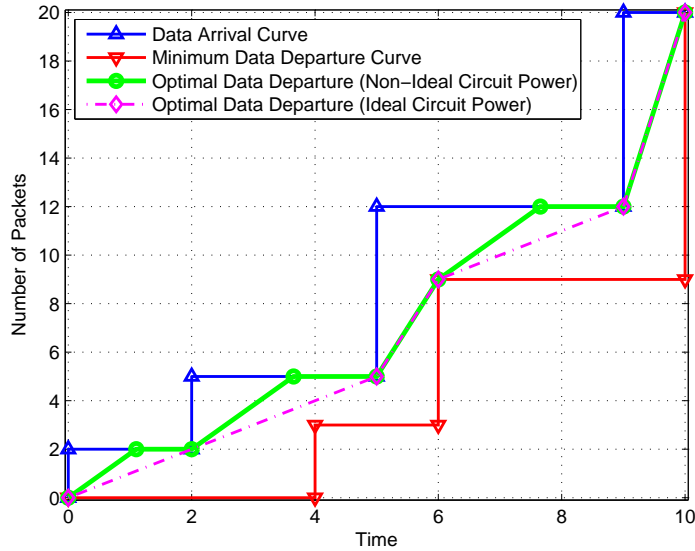}
\vspace{-0.8cm}
\caption{Data arrival, minimum data departure, and optimal data departure curves (ideal and non-ideal circuit power cases).}\label{visualization}
\vspace{-0.3cm}
\end{figure}

The optimal policy obtained by Algorithm 1 can be visualized by modifying the calculus approach in \cite{Zafer&Modiano2005}. Define the data arrival and minimum departure curves $A(t)$ and $D_{\text{min}}(t)$ as:
\begin{equation}\label{eq15}
\begin{array}{c}
\displaystyle A(t)=\sum_{i=0}^{A-1}{[a_i u(t-t_{\alpha_i})]}, \quad 0 \leq t \leq T,\\
\displaystyle D_{\text{min}}(t)=\sum_{j=1}^{D}{[d_j u(t-t_{\delta_j})]}, \quad 0 \leq t \leq T,
\end{array}
\end{equation}
where $u(t)$ is the unit-step function: $u(t)=1$, if $t \geq 0$, and $u(t)=0$ otherwise. In the ideal ($\rho=\beta=0$) circuit power case, the optimal transmission policy requires the transmitter to be always on, i.e. $l_n^{on*}=L_n$, $\forall{n}$ \cite{Chen08, Wang13}. In this case, consider a piece-wise linear data departure curve:
\begin{equation}\label{eq16}
\displaystyle D(t)=\sum_{m=1}^{n-1}{(r_m L_m)}+r_n(t-t_{n-1}), \;\; t_{n-1} \leq t \leq t_{n}, \;\; \forall{n},
\end{equation}
where the rates $r_n$ per epoch serve as the piece-wise slopes for $D(t)$. Following \cite{Zafer&Modiano2005}, the optimal departure curve $D^*(t)$ is shown to be the trajectory of letting a string tie its one end at the origin $(0, 0)$, pass the other end through $(T, G)$, and then taut between $A(t)$ and $D_{\text{min}}(t)$; see Fig.~\ref{visualization}. Consequently, the optimal $\tilde{r}_n^*$ for the ideal circuit power case can be derived from $D^*(t)$.

Interestingly, the optimal $\{r_n^*,l_n^{on*}\}$ for the non-ideal circuit power case can be simply obtained by an $r_{ee}$-clipping process over $\tilde{r}_n^*$. Specifically, we can set:
\begin{equation}\label{eq17}
\left\{
\begin{array}{lll}
r_n^*=r_{ee}, &l_n^{on*}=\frac{\tilde{r}_n^* L_i}{r_{ee}}, &\text{if} \quad \tilde{r}_n^*<r_{ee};\\
\displaystyle r_n^*=\tilde{r}_n^*, &l_n^{on*}=L_n, &\text{if} \quad \tilde{r}_n^* \geq r_{ee}.
\end{array}
\right.
\end{equation}
With \eqref{eq17}, the corresponding optimal data departure curve under non-ideal circuit power $D_{\alpha}^*(t)$ is plotted in Fig.~\ref{visualization}. Comparing to $D^*(t)$, the same amount of data $\Phi_n := \tilde{r}_n^* L_i$ departs per epoch $n$ in $D_{\alpha}^*(t)$. Yet, different from $D^*(t)$, an on-off transmission strategy is adopted when $\tilde{r}_n^* < r_{ee}$ for epoch $n$. This is because the total energy cost for $\Phi_n$ over such an epoch is in fact minimized by a transmission with $r_{ee}$ over an ``on'' period of length $l_n^{on*} = \Phi_n/r_{ee}< L_n$, i.e.,
\begin{equation}\label{eq.reward}
(P(r_{ee})+\rho)l_n^{on*}  = \frac{(P(r_{ee})+\rho)\Phi_n}{r_{ee}} = \Phi_n \min_{r \geq 0} \frac{P(r)+\rho}{r}   = \min_{r l_n^{on} =\Phi_n} (P(r)+\rho) l_n^{on}.
\end{equation}
For the epoches with $\Phi_n = \tilde{r}_n^* L_n \geq r_{ee} L_n$, however, any on-off strategy $(r_n, l_n^{on})$ with $r_n >\tilde{r}_n^*$ and $r_n l_n^{on}= \Phi_n$ only increases the energy cost since
\begin{equation}
(P(r_n)+\rho)l_n^{on} =\Phi_n \frac{P(r_n)+\rho}{r_n} > \Phi_n \frac{P(\tilde{r}_n^*)+\rho}{\tilde{r}_n^*}, 
\end{equation}
where the inequality is due to the fact that $\frac{P(r)+\rho}{r}$ is strictly increasing when $r \geq r_{ee}$. Hence, the data departures over these epoches remain the same in $D^*(t)$ and $D_{\alpha}^*(t)$.

It is worth noting that the optimal transmission strategy is in fact not unique in the on-off transmission epoches. In an on-off period including e.g., epoch $n_1$ to epoch $n_2$, different from $\{l_n^{on*}\}$ computed in (\ref{eq17}), we can have another set of $\{\bar{l}_n^{on*}\}$ such that $\sum_{n=n_1}^{n_2}\bar{l}_n^{on*} = \sum_{n=n_1}^{n_2}l_n^{on*}$. As long as $\{\bar{l}_n^{on*}\}$ are feasible, they are also optimal. In fact, we may even let $\bar{l}_n^{on*} =0$ (i.e., transmitter is ``off'') for some epoches, while carrying out transmission only over the remaining epoches in an on-off period per Lemma 1. 

\subsection{Development of Online Scheme}

To obtain the optimal benchmark, we assumed a non-causal case where complete information about the packet arrivals during the time interval $[0,T]$ is available. When a-priori knowledge of the future packet arrivals is not available in practice, we can develop a heuristic online scheme based on the proposed optimal offline policy. The idea is to schedule the packet transmissions according to the optimal rate control policy based on the current packet arrivals, and reschedule when new packets arrive. For instance, suppose that $a_0$ packets arrive at time instant $0$ with (different) deadline requirements. We can construct the set $\mathcal{D}$ in accordance to the deadline requirements, and let the set $\mathcal{A}=\{ (0,a_0),(\alpha_1,0)\}$, where $t_{\alpha_1}$ is determined by the largest deadline $t_{\delta_D}$. With such a $(\mathcal{A}, \mathcal{D})$, we run the proposed Algorithm~1 to find the optimal transmission policy until new packets arrive at $t$. Then we treat the current time instant $t$ as new ``$0$'' instant, and update the set $\mathcal{D}$ by subtracting all $t_{\delta_j}$ by $t$, removing the past deadlines (i.e. with negative $t_{\delta_j}$ after subtraction), and then including the deadline requirements for the newly arriving packets. The set $\mathcal{A}$ also needs to be updated. Note that we always have $\mathcal{A}=\{ (0,a_0),(\alpha_1,0)\}$, where $a_0$ is updated as the sum of the remaining packets in the buffer and the newly arriving packets, and $t_{\alpha_1}$ is determined by the last deadline in the updated $\mathcal{D}$. Algorithm~1 is run for the new $(\mathcal{A}, \mathcal{D})$, and the subsequent packet transmissions follow the resultant new policy. 
This process continues until all the packets are delivered.


\section{Generalization to Time-Varying Channel}
In this section, we generalize the proposed approach to a time-varying wireless channel, where the channel state $h$ in general changes with time. With a little abuse of notation, here we redefine an epoch as the interval between two consecutive channel or data state changing instants. Again, over the entire transmission interval $[0, T]$, assume that there are $N+1$ (channel or data) state changing instants $0=t_0<t_1<\cdots<t_N=T$. There are $N$ epoches with length $L_1:=t_1-t_0, \ldots, L_N:=t_N-t_{N-1}$. The packet arrival process and deadline requirements are modeled by set $\mathcal{A}:=\{(\alpha_i, a_i), i=0, \ldots, A\}$ and set $\mathcal{D}:=\{(\delta_j, d_j), j=1, \ldots, D\}$. Let $h_n$ denote the channel coefficient at epoch $n$; and $\mathcal{H}:=\{h_1, \ldots, h_N\}$.

With the power function $P(r_n; h_n):=\frac{1}{|h_n|^2}(e^{r_n}-1)$, we formulate the total energy consumption minimization problem over time-varying channels as follows:
\begin{equation}\label{tv.eq3}
\begin{array}{cll}
\displaystyle \min_{\boldsymbol{r}, \; \boldsymbol{l}^{on}} & \displaystyle \sum_{n=1}^{N}{[(P(r_n; h_n)+ \rho) l_n^{on}]} \\
\text{s.t.} &  \displaystyle \sum_{n=1}^{{\alpha_i}}{(r_n l_n^{on})} \leq \sum_{k=0}^{i-1}{a_k}, & i=1, \ldots, A, \\
& \displaystyle \sum_{n=1}^{{\delta_j}}{(r_n l_n^{on})} \geq \sum_{k=1}^{j}{d_k}, & j=1, \ldots, D, \\
& \displaystyle r_n \geq 0, \;\; 0 \leq l_n^{on} \leq L_n, & n=1, \ldots, N.
\end{array}
\end{equation}
%
%



With a change of variable $\Phi_n:=r_n l_n^{on}$, the non-convex problem in \eqref{tv.eq3} can be also reformulated into a convex program for $\{\Phi_n,l_n^{on}\}$. Let $\boldsymbol{\Lambda}^*:=\{\lambda_i^*,i=1,\ldots,A,\;\mu_j^*,j=1,\ldots,D\}$ collect the optimal Lagrange multipliers, and  $w_n:=\sum_{j=j_n}^{D}{\mu_j^*}-\sum_{i=i_n}^{A}{\lambda_i^*}$. Relying on the KKT conditions for the convex reformulation, we can follow the similar lines in Section III-A to derive
the sufficient and necessary optimality conditions for \eqref{tv.eq3}: 
\begin{equation}\label{tv.eq9}
\left.
\begin{split}
(r_n^*, l_n^{on*})=\arg \min \; & {[P(r_n; h_n)+ \rho -w_n r_n] l_n^{on}} \\
\text{s.t.} \quad & r_n \geq 0, \;\; 0 \leq l_n^{on} \leq L_n; \\
\end{split}
\right.
\;\; \forall n.
\end{equation}
\begin{equation}\label{tv.eq10}
\left \{
\begin{array}{c}
\lambda_i^*=0, \text{ if } \sum_{n=1}^{{\alpha_i}}{(r_n^* l_n^{on*})} < \sum_{k=0}^{i-1}{a_k},\\
\sum_{n=1}^{{\alpha_i}}{(r_n^* l_n^{on*})}=\sum_{k=0}^{i-1}{a_k}, \text{ if } \lambda_i^* > 0;
\end{array}
\right.
\;\; i=1, \ldots, A.
\end{equation}
\begin{equation}\label{tv.eq11}
\left \{
\begin{array}{c}
\mu_j^*=0, \text{ if } \sum_{n=1}^{{\delta_j}}{(r_n^* l_n^{on*})} > \sum_{k=1}^{j}{d_k},\\
\sum_{n=1}^{{\delta_j}}{(r_n^* l_n^{on*})} = \sum_{k=1}^{j}{d_k}, \text{ if } \mu_j^* > 0;
\end{array}
\right.
\;\; j=1, \ldots, D.
\end{equation}
%
%


For any $l_n^{on}>0$, we can derive from \eqref{tv.eq9} that
\begin{equation}\label{tv.eq12}
r_n^*=\arg \min_{r_n \geq 0}{[P(r_n; h_n)+\rho-w_n r_n]}.
\end{equation}
Let $P'(r_n; h_n)$ denote the first derivative of $P(r_n; h_n)$. We clearly have: $P'(r_n^*; h_n)=w_n$, leading to $r_n^* = \max\{0,\log(|h_n|^2 w_n)\}$, and $P(r_n^*; h_n)=\max\{0,w_n-\frac{1}{|h_n|^2}\}$. The latter is the celebrated water-filling form, where $w_n$ serves as a water-level.

Substituting $w_n=P'(r_n^*; h_n)$ into \eqref{tv.eq9} implies:
\begin{equation}\label{tv.eq13}
l_n^{on*}=\arg \min_{0 \leq l_n^{on} \leq L_n}{[P(r_n^*; l_n)+\rho-P'(r_n^*; h_n) r_n^*] l_n^{on}}.
\end{equation}
For each $h_n$ per epoch, we can obtain an EE-maximizing rate in \eqref{eq14}.
%
%
%
%
Note that $r_{ee}(h_n)$ is different for different $h_n$ per epoch. As with Lemma~1, relying on \eqref{tv.eq13}, we can show that:

\begin{lemma}
{\it The optimal transmission policy for (\ref{tv.eq3}) can only adopt one of the following three (``off'', ``on-off'' and ``on'') strategies per epoch $n$: (i) $l_n^{on*}=0$, (ii) $r_n^{*}=r_{ee}(h_n)$, $l_n^{on*} \leq L_n$, or (iii) $r_n^*>r_{ee}(h_n)$, $l_n^{on*}=L_n$.}
\end{lemma}
%


As with Lemma 2, we can establish that:

\begin{lemma}
{\it In the optimal policy for (\ref{tv.eq3}), the rates for epoches $n$ with $l_n^{on*}>0$ are given by: $r_n^*=P'^{-1}(w_n; h_n)$, where the water-level $w_n$ can only increase after a $t_{\alpha_i}$ where $\sum_{n=1}^{{\alpha_i}}{(r_n^* l_n^{on*})}=\sum_{k=0}^{i-1}{a_k}$, and decrease after a $t_{\delta_j}$ where $\sum_{n=1}^{{\delta_j}}{(r_n^* l_n^{on*})}=\sum_{k=1}^{j}{d_k}$.}
\end{lemma}

Similar to the time-invariant channel case, Lemma 3 states that the optimal policy depends on the EE-maximizing rates; i.e., any transmit-rate less than the EE-maximizing rate $r_{ee}(h_n)$ should not be adopted at epoch $n$. However, the value of $r_{ee}(h_n)$ is in general different for different $h_n$ across epoches. In both time-invariant and time-varying channel cases, the change of ``water-level'' $w_n$ follows the same pattern: it increases after a casuality constraint becomes tight, and decreases after a deadline constraint is tight. Given the same $w_n$, the same transmit-rate is maintained in the time-invariant case. This leads to the $r_{ee}$-clipped ``string-tautening'' procedure in Algorithm 1. For the time-varying channel case, the same water-level yields different transmit-power and rate when channel $h_n$ varies; specifically, higher rates are employed for better channels through a water-filling type power allocation for the most efficient energy usage.
This revealed structure of the optimal policy implies that we can modify Algorithm 1 to implement a water-level based ``string-tautening'' approach to finding the optimal solution for \eqref{tv.eq3}.

\begin{figure}[b]
\vspace{-0.3cm}
\centering
\includegraphics[width=4in]{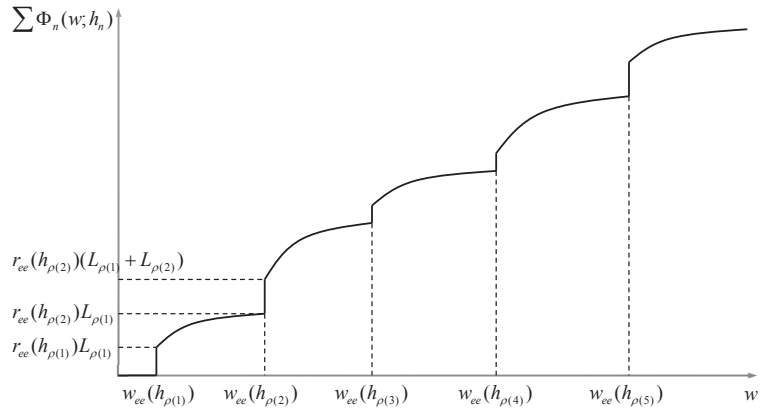}
\vspace{-0.6cm}
\caption{Data departure for a given water-level $w$: Suppose a transmission over 5 epoches. Sort $w_{ee}(h_n)$ in ascending order such that: $w_{ee}(h_{\rho(1)}) < \cdots < w_{ee}(h_{\rho(5)})$. (i) If $w<w_{ee}(h_{\rho(1)})$, it follows from (\ref{tv.eq17}) that $\Phi_n(w; h_n)=0$, $\forall n$; hence, the total data departure is 0. (ii) When $w=w_{ee}(h_{\rho(1)})$, $\Phi_n(w; h_n)=0$, $\forall n \neq \rho(1)$; the total departure is given by $\Phi_n(w; h_{\rho(1)})$ which ranges from 0 to $r_{ee}(h_{\rho(1)}) L_{\rho(1)}$, depending on the value of $l_{\rho(1)}^{on}\leq  L_{\rho(1)}$. (iii) For $w_{ee}(h_{\rho(1)})<w<w_{ee}(h_{\rho(2)})$, the total departure is still $\Phi_n(w; h_{\rho(1)})$ which increases as $w$ increases. (iv) When $w=w_{ee}(h_{\rho(2)})$, $\Phi_n(w; h_n)=0$, $\forall n \neq \rho(1), \rho(2)$; the total departure is $\Phi_n(w; h_{\rho(1)}) + \Phi_n(w; h_{\rho(2)})$, which ranges from $r_{ee}(h_{\rho(2)}) L_{\rho(1)}$ to $r_{ee}(h_{\rho(2)}) (L_{\rho(1)} +L_{\rho(2)})$. And so on.}\label{phi}
\vspace{-0.5cm}
\end{figure}

To this end, let $w_{\alpha_i}^+$ or $w_{\delta_j}^-$ denote the constant water-level to make the $i$th causality or the $j$th deadline constraint become tight at $t_{\alpha_i}$ or $t_{\delta_j}$.
Given an invariant water-level $w$ before $t_{\alpha_i}$ (or $t_{\delta_j}$), the optimal rate per epoch $n$ is given by $r_n^*=P'^{-1}(w; h_n)$ if $l_n^{on*}>0$. Define:
\begin{equation}\label{tv.eq15}
w_{ee}(h_n):=P'(r_{ee}(h_n); h_n), \quad \forall{n}.
\end{equation}
Since we must have $r_n^* \geq r_{ee}(h_n)$ if $l_n^{on*}>0$ per Lemma~3, we can have $l_n^{on*}>0$ only when $w \geq w_{ee}(h_n)$ (recall that $P'()$ is an increasing function). With the water-level $w$, the optimal strategy per epoch $n$ is then:
\begin{equation}\label{tv.eq16}
\left\{
\begin{array}{ll}
l_n^{on*}=0, & \text{if} \; w<w_{ee}(h_n), \\
r_n^*=r_{ee}(h_n), \; l_n^{on*} \leq L_n, & \text{if} \; w=w_{ee}(h_n), \\
r_n^*=P'^{-1}(w; h_n), \; l_n^{on*}=L_n, & \text{if} \; w>w_{ee}(h_n).
\end{array}
\right.
\end{equation}
Define data departure $\Phi_n(w; h_n)=r_n^* l_n^{on*}$ per epoch $n$. By \eqref{tv.eq16}, we have:
\begin{equation}\label{tv.eq17}
\left\{
\begin{array}{ll}
\Phi_n(w; h_n)=0, & \text{if} \; w<w_{ee}(h_n), \\
\Phi_n(w; h_n) \in [0, r_{ee}(h_n) L_n], & \text{if} \; w=w_{ee}(h_n), \\
\Phi_n(w; h_n)=P'^{-1}(w; h_n) L_n, & \text{if} \; w>w_{ee}(h_n).
\end{array}
\right.
\end{equation}
Using (\ref{tv.eq17}), the values of $w_{\alpha_i}^+$ and $w_{\delta_j}^-$ can be calculated by solving the equations:
\begin{equation}\label{tv.eq18}
\displaystyle \sum_{n=1}^{{\alpha_i}}{\Phi_n(w_{\alpha_i}^+; h_n)=\sum_{k=0}^{i-1}{a_k}}, \;\; i=1, \ldots, A; \qquad   \displaystyle \sum_{n=1}^{{\delta_j}}{\Phi_n(w_{\delta_j}^-; h_n)=\sum_{k=1}^{j}{d_k}}, \;\; j=1, \ldots, D.
\end{equation}
Note that $\sum_{n=1}^{{\alpha_i}}{\Phi_n(w; h_n)}$ and $\sum_{n=1}^{{\delta_j}}{\Phi_n(w; h_n)}$ are increasing in $w$; see an example in Fig.~\ref{phi}. 
As a result, the equations in \eqref{tv.eq18} can be solved by a bisectional search.

With $w_{{\alpha_i}}^+$ and $w_{{\delta_j}}^-$ obtained, the optimal $\{ \boldsymbol{r}^*, \boldsymbol{l}^{on*} \}$ for \eqref{tv.eq3} can be computed by a $w_{ee}(h_n)$-clipped ``water-tautening'' approach in Algorithm~2.

%
%
%
%
%
%
\begin{spacing}{1.13}
\small
\vspace{0.25 in}
\hrule height 0.1pt depth 0.3pt
\vspace{0.05 in}
\noindent {\bf Algorithm~2} $w_{ee}(h_n)$-Clipped ``Water-Tautening''
\vspace{0.05 in}
\hrule height 0.1pt depth 0.3pt
\vspace{0.1 in}
\begin{algorithmic}[1]
\Procedure {ScheduleW}{${\cal{A}, \cal{D}, \cal{H}}$}
	\State $N_{\text{offset}}=0$, $r_n^*=0$, $l_n^{on*}=0$, $\forall{n}$;
	\While {$N_{\text{offset}}<N$}
		\State [$\tau$, $w$, $\Delta$]=FirstChange($\cal{A}$, $\cal{D}$, $\cal{H}$);
		\For {$n=1$ to ${\tau}$}
			\State $r_{N_{\text{offset}}+n}^*=P'^{-1}(w;h_n)$;
			\If {${w>w_{ee}(h_n)}$}
				\State $l_{N_{\text{offset}}+n}^{on*}=L_n$;
			\EndIf
		\EndFor
		\If {there exists $n_{ee}$ with $w_{ee}(h_{n_{ee}})=w$}
            \State $l_{N_{\text{offset}}+n_{ee}}^{on*}=\frac{\Delta-\sum_{n=1, n \neq n_{ee}}^{n_{\tau}}{(r_{N_{\text{offset}}+n}^* l_{N_{\text{offset}}+n}^{on*}})}{r_{ee}(h_{n_{ee}})}$;
		\EndIf
		\State $N_{\text{offset}}=N_{\text{offset}}+{\tau}$, and update $\cal{A}, \cal{D}, \cal{H}$;
	\EndWhile
\EndProcedure
\Statex
\Function {[$\tau$, $w$, $\Delta$]=FirstChangeW}{$\cal{A}$, $\cal{D}$, $\cal{H}$}
	\State $w^{-}=0$, $w^{+}=\infty$, $\tau^{-}=\tau^{+}=0$, $\Delta^- =\Delta^+ =0$;
	\State sort $\alpha_i$, $\delta_j$ in ascending order into $\boldsymbol{\pi}:=\{ \pi_1, \ldots, \pi_{A+D}\}$;
	\For {$k=1$ to $A+D$}
		\If {$\pi_k=\alpha_i \in \boldsymbol{\alpha}$ for a certain $i$}
            \State calculate $w_{{\alpha_i}}^+$ by solving (\ref{tv.eq18});
            \If {$w_{{\alpha_i}}^+ \leq w^{+}$}
                \State $\tau^{+}=\alpha_i$, $w^{+}=w_{{\alpha_i}}^+$, $\Delta^+=\sum_{k=0}^{i-1}{a_k}$;
            \EndIf
		\ElsIf {$\pi_k=\delta_j \in \boldsymbol{\delta}$ for a certain $j$}
            \State calculate $w_{{\delta_j}}^-$ by solving (\ref{tv.eq18});
            \If {$w_{{\delta_j}}^- \geq w^{-}$}
                \State $\tau^{-}=\delta_j$, $w^{-}=w_{{\delta_j}}^-$, $\Delta^-=\sum_{k=1}^{j}{d_k}$;
            \EndIf
		\EndIf
		\If {$w^{-}>w^{+}$ \& $\tau^{-}<\tau^{+}$}
			\State return $\tau=\tau^{-}$, $w=w^{-}$, $\Delta = \Delta^-$;
		\ElsIf {$w^{-} \geq w^{+}$ \& $\tau^{-} \geq \tau^{+}$}
			\State return $\tau=\tau^{+}$, $w=w^{+}$, $\Delta = \Delta^+$;
		\EndIf
	\EndFor
\EndFunction
\end{algorithmic}
\vspace{0.05 in}
\hrule height 0.1pt depth 0.3pt
\vspace{0.25 in}
\end{spacing}
\vspace{0.2 in}

The key component in Algorithm~2 is the function FirstChangeW, which determines the first water-level changing time $t_{\tau}$ and the water-level $w$ used before $t_{\tau}$. The two candidate water-levels are updated as: $w^+=\min_{\alpha_i \leq {n}}{w_{{\alpha_i}}^+}$ and $w^-=\max_{\delta_j \leq {n}}{w_{{\delta_j}}^-}$, which are in fact the maximum and minimum values for an invariant water-level to satisfy all the causality and deadline constraints before $t_n$, respectively; and $\tau^+$, $\tau^-$ are the corresponding $\alpha_i$ or $\delta_j$ yielding $w^+$, $w^-$. If we have $w^+<w^-$ at a certain $t_{n}$, then the water-level needs to be changed before $t_{n}$ since no invariant water-level can satisfy all the causality and deadline constraints so far. The first water-level changing time can be obtained by comparing $\tau^+$ and $\tau^-$ to see which type of constraint first becomes tight. When the returned $t_{\tau}<T$, Function FirstChangeW can be reused for a new $(\cal{A}, \cal{D}, \cal{H})$ system over the remaining time to find the next water-level changing time and the next water-level. Note that Function FirstChangeW also returns the delivered data amount $\Delta$ before $t_\tau$, which is necessary for determining the length of the ``on'' period for the possible on-off epoch $n_{ee}$ with $w_{ee}(h_{n_{ee}})=w$.

As with Theorem 1, we can establish that:

\begin{theorem}
{\it Algorithm 2 computes the optimal transmission policy for \eqref{tv.eq3} with a linear complexity ${\cal O}(A+D)$.}
\end{theorem}

Again, we prove Theorem~2 by showing the existence of a Lagrange multiplier vector $\boldsymbol{\Lambda}^*$, with which $\boldsymbol{r}^*$ and $\boldsymbol{l}^{on*}$ satisfy the sufficient and necessary optimality conditions \eqref{tv.eq9}--\eqref{tv.eq11}. In the search of rate-changing points, we only need to go through the $A+D$ data arrival or deadline time instants.


\section{Numerical Results}

%
%
\begin{figure}[t]
\vspace{-0.3cm}
\centering
\includegraphics[width=4.2in]{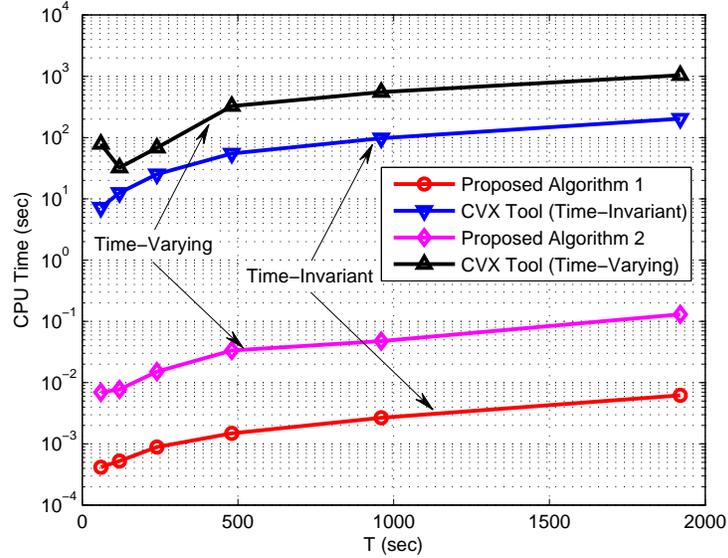}
\vspace{-0.8cm}
\caption{Comparison of average CPU time for the proposed algorithms and CVX toolbox.}\label{cputime}
\vspace{-0.3cm}
\end{figure}
%
%
Consider packet transmission over $T$ seconds and the number of total packet arrivals is $G=40$. The bandwidth for the system is $1$ KHz while each packet contains $1$K bits. The power-rate relationship is dictated by the Shannon-capacity formula in \eqref{eq2}. The non-zero circuit power during the ``on'' mode is $\rho=3$ Watts.
(i) Assume first a time-invariant channel with power gain $|h|^2=2$ and unit noise variance. To validate the proposed Algorithm~1, we use it and the standard CVX toolbox to solve \eqref{eq3} when the transmission interval $T=60$, $120$, $240$, $480$, $960$, $1920$ seconds. For each $T$ value, we test $50$ trial cases where the intervals $\{t_{\alpha_i}-t_{\alpha_{i-1}}\}$ and $\{t_{\delta_j}-t_{\delta_{j-1}}\}$ follow a uniform distribution with mean $T/10$, respectively. It is confirmed that the two methods yield the same rate control policies for all trial cases, demonstrating the correctness of the proposed Algorithm~1. On the other hand, since Algorithm~1 directly constructs the optimal solution for the problem at hand relying on the optimality conditions, it is much more efficient than the CVX toolbox in terms of computational complexity. Fig. \ref{cputime} depicts the average CPU time required for the two methods. 
We can observe that the required CPU time of the proposed Algorithm~1 is less than 0.01\% of that of the standard CVX toolbox for all $T$ values.
(ii) Consider next a time-varying channel where the random channel coefficients per second are independently generated from a Rayleigh distribution with average power $|\bar h|^2=2$. We use Algorithm 2 and the standard CVX toolbox to solve \eqref{tv.eq3}. Again, it is confirmed that Algorithm~2 and the CVX toolbox yield the same rate control policies for all trial cases. Fig. \ref{cputime} also includes the average required CPU time with these two methods. It is observed that Algorithm~2 only requires a CPU time less than 0.1\% of that with the CVX toolbox for all $T$ values. The significantly reduced complexity can clearly benefit the real-time implementation of the algorithm in e.g., the proposed online scheme.
\begin{figure}[t]
\vspace{-0.3cm}
\centering
\includegraphics[width=4.2in]{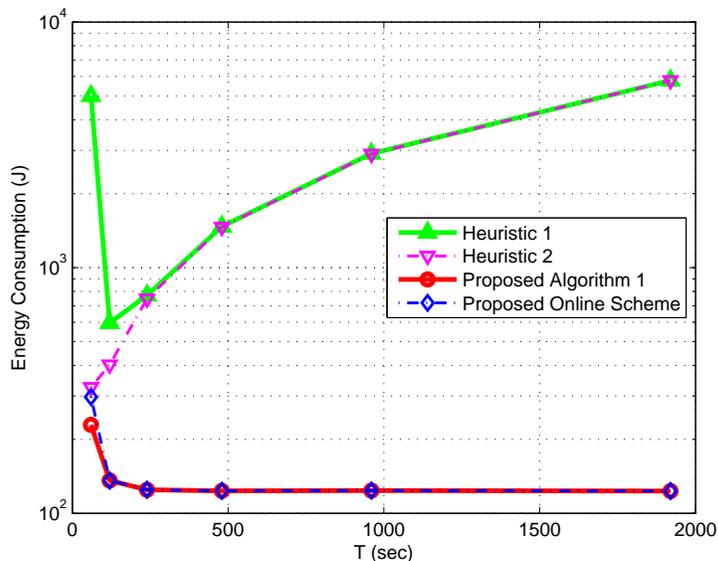}
\vspace{-0.8cm}
\caption{Comparison of average energy consumption for Algorithm~1, proposed online scheme, and heuristic approaches.}\label{energy}
\vspace{-0.3cm}
\end{figure}

Next, 
in terms of energy consumption, we compare the optimal policies obtained by our proposed Algorithm~1 and its online scheme (as described in Section~III-D), as well as two heuristic policies in the time-invariant channel case.
Heuristic~1 is obtained by always selecting a rate to meet the next active causality or deadline constraint with equality. Heuristic~2 is obtained by using the calculus approach in \cite{Zafer&Modiano2005} under an ideal circuit-power assumption, even though the circuit-power consumption is in fact non-ideal. Both heuristic approaches can provide a feasible policy that satisfies all the causality and deadline constraints in \eqref{eq3} for fair comparison. Fig. \ref{energy} provides the average energy consumption with the policies obtained by Algorithm~1 and the two heuristic approaches, when $G=40$, and $T=60\sim1920$ seconds. Again, for each $T$ value, we test 50 trial cases where the causality and deadline constraints are randomly generated. The proposed Algorithm~1 significantly outperforms the two heuristic counterparts, always yielding the smallest energy consumption for every $T$ value. Compared to the heuristic policies, the energy saved from the optimal policy becomes more significant as the value of $T$ increases, i.e., as the entire transmission interval becomes longer. For small $T$ values, Heuristic~2 has approximately the same performance as the proposed Algorithm~1. The reason is that the constraints are usually strict and thus the optimal transmit-rates are usually larger than $r_{ee}$ when $T$ value is small; as a result, in the optimal transmission policy produced by Algorithm~1, there are very few on-off epoches, i.e. the transmitter is always selected to be on, rendering the policies by Algorithm~1 and Heuristic~2 become almost identical. For large $T$ values, Heuristic~2 scheme becomes almost the same as the naive Heuristic~1. The optimal policy can save up to 100 times the energy over the naive Heuristic~1 for small $T$, and can save $10\sim100$ times the energy over the heuristic ones for large $T$. It is also observed that the energy consumption of the proposed online scheme is close to that of the optimal offline strategy, especially for large $T$. In such a case, even the proposed online scheme without the knowledge of future arrivals can save almost $10\sim100$ times the energy over the heuristic ones with complete a-priori information about the packet arrivals. 

%
%
%
%

\begin{figure}[t]
\vspace{-0.3cm}
\centering
\includegraphics[width=4.2in]{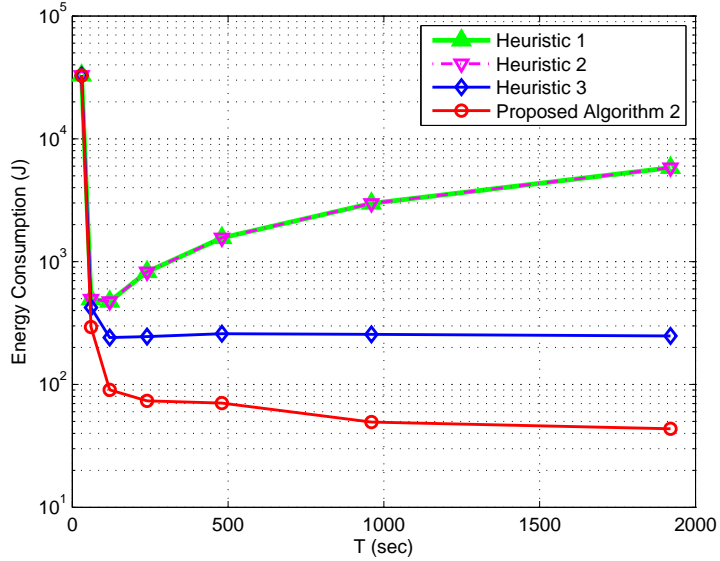}
\vspace{-0.8cm}
\caption{Comparison of average energy consumption for Algorithm~2 and heuristic approaches.}\label{energy_tv}
\vspace{-0.3cm}
\end{figure}

Lastly, we compare the energy consumption between the proposed Algorithm~2, Heuristic~1 and 2, as well as another heuristic method (called Heuristic~3) in the time-varying channel case.
Specifically, Heuristic~3 is obtained by using Algorithm~1 under a time invariant channel assumption, even though the channel is in fact time-varying.  Fig.~\ref{energy_tv} shows the energy consumption of these schemes, where each result is averaged over 50 random trials. The proposed Algorithm~2 apparently outperforms the heuristic counterparts, always yielding the smallest energy consumption for each trial case. It is observed that the optimal policy obtained by our Algorithm~2 can save up to 100 times the energy over Heuristic~1 and 2, and 5 times the energy over Heuristic~3, especially for large $T$ values.
%
%


\section{Conclusion}

We proposed a novel unified approach to optimizing the energy-efficient transmission policy for delay-limited bursty packet arrivals under non-ideal circuit power consumption. Efficient algorithms were developed to find the optimal offline solutions with a low computational complexity for both time-invariant and time-varying channels. An insightful visualization was presented to reveal the specific structure of the optimal policy. Our approach can provide the optimal benchmarks for practical schemes. Based on the proposed optimal offline policies, development of energy-efficient online scheduling policies was also discussed and will be further explored in our future work. Generalization of our approach to wireless networks with multiple interfering links will be also pursued.


\appendix
\subsection{Proof of Lemma~1}
Define $\xi_{ee}(r):=\frac{P(r)+\rho}{r}$. Taking the first derivative of $\xi_{ee}(r)$, we have:
\begin{equation}\label{eqa2}
\displaystyle \frac{d\xi_{ee}(r)}{dr}=\frac{P'(r)r-(P(r)+\rho)}{r^2}.
\end{equation}
Due to its ``convex-over-linear'' form, we can show that $\xi_{ee}(r)$ first decreases and then increases with $r$, and it reaches the minimum at $r_{ee}$. This implies:
\begin{equation}\label{eqa3}
\left\{
\begin{array}{ll}
\displaystyle P'(r)r-(P(r)+\rho)<0, &\text{if} \quad r<r_{ee},\\
\displaystyle P'(r)r-(P(r)+\rho)=0, &\text{if} \quad r=r_{ee},\\
\displaystyle P'(r)r-(P(r)+\rho)>0, &\text{if} \quad r>r_{ee}.
\end{array}
\right.
\end{equation}
If we have an $r_n^*<r_{ee}$ when $l_n^{on*}>0$, it follows from \eqref{eqa3} that $P'(r_n^*)r_n^*-(P(r_n^*)+\rho)<0$. But when $P'(r_n^*)r_n^*-(P(r_n^*)+\rho)<0$, \eqref{eq13} implies that $l_n^{on*}=0$, which leads to a contradiction. Hence, $r_n^*<r_{ee}$ is not allowed when $l_n^{on*}>0$.

When $r_n^*>r_{ee}$, we have $P'(r_n^*)r_n^*-(P(r_n^*)+\rho)>0$ according to \eqref{eqa3}. This together with \eqref{eq13} then dictates $l_n^{on*}=L_n$. In the case of $r_n^*=r_{ee}$, we have $P'(r_n^*)r_n^*-(P(r_n^*)+\rho)=0$, so any $l_n^{on*} \in{[0,L_n]}$ is a minimizer in \eqref{eq13}.

\subsection{Proof of Lemma~2}

Clearly, $r_n^*=P'^{-1}(w_n)$ changes only when $w_n$ changes its value. By the definition $w_n=\sum_{j=j_n}^{D}{\mu_j^*}-\sum_{i=i_n}^{A}{\lambda_i^*}$,
if $\lambda_i^* = 0$, $\forall i = 1, \ldots, A-1$, and $\mu_j^*=0$, $\forall j = 1, \ldots, D-1$, then a constant $w=\mu_D^* - \lambda_A^*$ will be used over all the epoches. We will have a change only when $\lambda_i^* >0$ for a certain $i \in [1,A-1]$, or $\mu_j^* >0$ for a certain $j \in [1,D-1]$, which occurs at the corresponding $t_{\alpha_i}$ or $t_{\delta_j}$. In addition, it follows from the complementary slackness conditions \eqref{eq10}--\eqref{eq11} that we must have $\sum_{n=1}^{\alpha_i}{(r_n^* l_n^{on*})}=\sum_{k=0}^{i-1}{a_k}$ or $\sum_{n=1}^{\delta_j} (r_n^* l_n^{on*}) = \sum_{k=1}^j d_k$ at such a $t_{\alpha_i}$ or $t_{\delta_j}$.

If a change occurs at a certain $t_{\alpha_i}$, then $\lambda_i^*>0$. For the epoch $n= \alpha_i$, we have $i_n = \arg \min_l \{l: n\leq \alpha_l\} = i$; thus $w_{\alpha_i} = \sum_{j=j_n}^D \mu_j^* - \sum_{l=i}^A \lambda_l^*$. On the other hand, for the epoch $n= \alpha_i+1$, we have $i_n = \arg \min_l \{l: n\leq \alpha_l\} = i+1$; thus $w_{\alpha_i+1} = \sum_{j=j_n}^D \mu_j^* - \sum_{l=i+1}^A \lambda_l^*$. Therefore, $w_{\alpha_i+1}-\omega_{\alpha_i} = \lambda_i^* >0$; consequently, the rate increases after $t_{\alpha_i}$ since $P'^{-1}(w_n)$ is an increasing function of $w_n$.

If a change occurs at a certain $\delta_j$, then $\mu_j^*>0$. For the epoches $n= \delta_j$ and $n= \delta_j+1$, we can similarly derive that $w_{\delta_j+1}-\omega_{\delta_j} = -\mu_j^* <0$; consequently, the rate decreases after $t_{\delta_j}$.


%

\subsection{Proof of Theorem~1}

\begin{figure}[t]
\vspace{-0.4cm}
\centering
\includegraphics[width=4in]{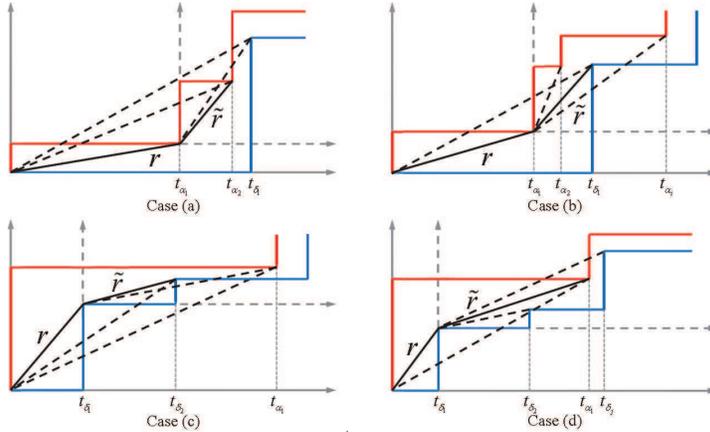}
\vspace{-0.6cm}
\caption{(i) If the first rate-changing point is found at $\tau = \tau^+ =\alpha_i$ by Function FirstChangeR, the rate used before $t_{\tau}$ is then given by $r=r_{\alpha_i}^+$, and the $i$th causality constraint is tight: $\sum_{n=1}^{\alpha_i} r l_n^{on*} = \sum_{k=0}^{i-1}a_k$. Selecting $\tau = \tau^+$ also implies there exists a $\tau^- =\delta_j > \tau$ such that $r_{\alpha_k}^+ > r$, $\forall \alpha_i < \alpha_k \leq \delta_j$, and $r_{\delta_j}^-> r$. Suppose w.l.o.g. that $\tau = \alpha_1$, and we have $\delta_1 > \alpha_1$ with $r_{\delta_1}^-> r$ and $\alpha_2 \in (\alpha_1, \delta_1)$ with $r_{\alpha_2}^+ > r$. After selecting $\alpha_1$ as the first rate-changing point, we construct the new $({\cal A}, {\cal D})$ system where $t_{\alpha_1}$ becomes 0 and $t_{\alpha_2}$ becomes the new $\tilde{t}_{\alpha_1}$ (we use $\tilde{}$ to distinguish the new system from the old one). Case (a): if $\tilde{r}^+_{\alpha_1} < \tilde{r}_{\delta_1}^-$, then Function FirstChangeR selects $\alpha_2$ as the next rate-changing point and rate $\tilde{r}=\tilde{r}_{\alpha_1}^+ > r_{\alpha_2}^+ > r$. Case (b): if $\tilde{r}^+_{\alpha_1} \geq \tilde{r}_{\delta_1}^-$, then Function FirstChangeR selects $\delta_1$ as the next rate-changing point (when there exists an $\alpha_i$ with $\tilde{r}^+_{\alpha_i} < \tilde{r}_{\delta_1}^-$) or a constant rate is maintained till the end (otherwise). In either situation, we have $\tilde{r}\geq \tilde{r}_{\delta_1}^- > r_{\delta_1}^- > r$. (ii) Similarly, if the rate $r$ is changed at $t_{\tau}$ where $\sum_{n=1}^{\tau}{r l_n^{on*}}=\sum_{k=1}^{j}{d_k}$, we can show that the next rate $\tilde{r}<r$; see Cases (c) and (d).} \label{appendix}
\vspace{-0.5cm}
\end{figure}

Due to the rules used in the function FirstChangeR, it can be shown that the rate-changing pattern in the transmission policy $\mathcal{R}$ produced by Algorithm~1 is consistent with the optimal structure revealed in Lemma~2, i.e., (i) if the rate in use is first $r$ and then changed to $\tilde{r}$ at $t_{\tau}$ where $\sum_{n=1}^{\tau}{r l_n^{on*}}=\sum_{k=0}^{i-1}{a_k}$, then we must have $\tilde{r}>r$; and (ii) if the rate $r$ is changed at $t_{\tau}$ where $\sum_{n=1}^{\tau}{r l_n^{on*}}=\sum_{k=1}^{j}{d_k}$, then we must have the next rate $\tilde{r}<r$. Fig.~\ref{appendix} provides an illustration and the sketch of the proof for this claim.

Suppose that the rate changes $M$ times in $\mathcal{R}:=\{ r_n^*, l_n^{on*}, n=1, \ldots, N \}$ yielded by Algorithm~1. We divide the policy into $M+1$ phases: rate $r_n^*=\check{r}_1$ over epoches $n \in [1, \tau_1]$, $r_n^*=\check{r}_2$ over epoches $n \in [\tau_1+1, \tau_2]$, $\ldots$, $r_n^*=\check{r}_{M+1}$ over epoches $n \in [\tau_{M}+1, N]$. We can then construct a set of Lagrange multipliers $\boldsymbol{\Lambda}^*:=\{ \lambda_i^*, i=1, \ldots, A, \mu_j^*, j=1, \ldots, D \}$ as follows:

Let $\mu_D^*=P'(\check{r}_{M+1})>0$, where the inequality is due to the strictly increasing of $P(r)$, leading to positivity of $P(r)$. Let $\lambda_{\tau_{m}}^*=P'(\check{r}_{m+1})-P'(\check{r}_{m})$, if  $\tau_m=\alpha_i \in \boldsymbol{\alpha}$ for a certain $i$ and $\sum_{n=1}^{\tau_m}{r_n^* l_n^{on*}}=\sum_{k=0}^{i-1}{a_k}$, or let $\mu_{\tau_m}^*=P'(\check{r}_{m})-P'(\check{r}_{m+1})$, if $\tau_m=\delta_j \in \boldsymbol{\delta}$ for a certain $j$ and $\sum_{n=1}^{\tau_m}{r_n^* l_n^{on*}}=\sum_{k=1}^{j}{d_k}$, $\forall m=1, \ldots, M$. We have shown that the rate $\check{r}_{m+1}>\check{r}_m$ if the causality constraint is tight at $t_{\tau_m}$, or $\check{r}_{m+1}<\check{r}_m$ if the deadline constraint is tight at $t_{\tau_m}$. Recalling that $P'(r)$ is increasing in $r$, it readily follows that $\lambda_{\tau_{m}}^*>0$ or $\mu_{\tau_m}^*>0$, depending on which type of constraint is tight at $t_{\tau_{m}}$. Except these $M+1$ positive $\mu_D^*$ and $\lambda_{\tau_{m}}^*$ or $\mu_{\tau_m}^*$, all other Lagrange multipliers in $\boldsymbol{\Lambda}^*$ are set to zero.

With such a $\boldsymbol{\Lambda}^*$, the complementary slackness conditions \eqref{eq10}--\eqref{eq11} clearly hold. Using such a $\boldsymbol{\Lambda}^*$ also leads to $w_n:=\sum_{j=j_n}^{D}{\mu_j^*}-\sum_{i=i_n}^{A}{\lambda_i^*}=P'(\check{r}_{m})$, $\forall n \in [\tau_{m-1}+1, \tau_m]$ (with $\tau_0:=1$ and $\tau_{M+1}:=N$). This implies that $r_n^*=\check{r}_m=\arg \min_{r_n \geq 0}{P(r_n)+\rho-w_n r_n}$, $\forall n \in [\tau_{m-1}+1, \tau_m]$. In addition, the construction of $\mathcal{R}$ ensures $l_n^{on*}=L_n$ when $r_n^*=\check{r}_m>r_{ee}$, and computes a feasible set of $l_n^{on*} \leq L_n$ when $r_n^*=\check{r}_m=r_{ee}$ in each phase $m$. This guarantees that each pair of $(r_n^*, l_n^{on*})$ satisfies \eqref{eq9}; thus, $\{ r_n^*, l_n^{on*}, n=1, \ldots, N \}$ follows the optimal structure in Lemma~1.

We have proven that $\{ \boldsymbol{r}^*, \boldsymbol{l}^{on*} \}$ yielded by Algorithm~1 and the Lagrange multipliers $\boldsymbol{\Lambda}^*$ constructed accordingly, satisfy the sufficient and necessary optimality conditions \eqref{eq9}--\eqref{eq11} for \eqref{eq3}.  It then readily follows that $\mathcal{R}$ is a global optimal policy for \eqref{eq4}. In the search of the rate-changing points and the associated rates in Algorithm~1, we only need to go through the $A+D$ data arrival or deadline time instants as shown in Fig. 2; hence, the algorithm has a complexity ${\cal O}(A+D)$.


\begin{thebibliography}{99}

\bibitem{ICC14} Z. Nan and X. Wang, ``Energy-efficient transmission of delay-limited bursty data packets under non-ideal circuit power consumption,'' in {\em Proc. ICC}, Sydney, Australia, Jun. 2014.

\bibitem{ChinaSIP14} Z. Nan, T. Chen and X. Wang, ``Energy-efficient transmission of delay-limited bursty data packets over time-varying channels under non-ideal circuit power,'' in {\em Proc. ChinaSIP}, Xi'an, China, July 2014.

\bibitem{Berry02} R. Berry and R. Gallager, ``Communication over fading channels with delay constraints,'' {\em IEEE Trans. Inf. Theory}, vol. 48, no. 5, pp. 1135--1149, May 2002.

\bibitem{Uysal04} E. Uysal-Biyikoglu and A. E. Gamel, ``On adaptive transmission for energy efficient in wireless data networks,'' {\em IEEE Trans. Inf. Theory}, vol. 50, no. 12, pp. 3081--3094, Dec. 2004.

\bibitem{ElGamal02} A. El Gamal, C. Nair, B. Prabhakar, E. Uysal-Biyikoglu, and S. Zahedi, ``Energy-efficient scheduling of packet transmissions over wireless networks,'' in {\em Proc. INFOCOM}, vol. 3, pp. 1773--1783, 2002.

\bibitem{Yao&Giannakis2005} Y. Yao and G. B. Giannakis, ``Energy-efficient scheduling for wireless sensor networks,'' {\em IEEE Trans. Commun.}, vol. 53, no. 8, pp. 1333--1342, Aug. 2005.

\bibitem{Wan08IT} X. Wang and G. Giannakis, ``Power-efficient resource allocation for time-division multiple access over fading channels,'' {\em IEEE Trans. Inf. Theory}, vol. 54, no. 3, pp. 1225--1240, Mar. 2008.



\bibitem{Uysal-Biyikoglu02net} E. Uysal-Biyikoglu, B. Prabhakar, and A. El Gamal, ``Energy-efficient packet transmission over a wireless link,'' {\em IEEE/ACM Trans. Netw.}, vol. 10, no. 4, pp. 487--499, Aug. 2002.

\bibitem{Zafer&Modiano2005} M. Zafer and E. Modiano, ``A calculus approach to minimum energy transmission policies with quality of service guarantees,'' in {\em Proc. INFOCOM}, vol. 1, pp. 548--559, 2005.

\bibitem{Chen08} W. Chen, M. Neely, and U. Mitra, ``Energy-efficient transmissions with individual packet delay constraints,'' {\em IEEE Trans. Inf. Theory}, vol. 54, no. 5, pp. 2090--2109, May 2008.

\bibitem{Chen09} W. Chen, U. Mitra, and M. Neely, ``Energy-efficient scheduling with individual packet delay constraints over a fading channel,'' {\em Wireless Netw.}, vol. 15, no. 5, pp. 601--618, 2009.

\bibitem{Wang13} X. Wang and Z. Li, ``Energy-efficient transmissions of bursty data packets with strict deadlines over time-varying wireless channels,'' {\em IEEE Trans. Wireless Commun.}, vol. 12, no. 5, pp. 2533--2543, May 2013.

\bibitem{Xu13} J. Xu and R. Zhang, ``Throughput optimal policies for energy harvesting wireless transmitters with non-ideal circuit power,'' to appear in {\em IEEE J. Sel. Areas Commun.}.

\bibitem{Bai11} Q. Bai, J. Li, and J. Nossek, ``Throughput maximizing transmission strategy of energy harvesting nodes,'' in {\em Proc. IWCLD}, 2011.

\bibitem{Orh12} O. Orhan, D. Gunduz, and E. Erkip, ``Throughput maximization for an energy harvesting system with processing cost,'' in {\em Proc. ITW}, 2012.

\bibitem{Blu10} O. Blume, H. Eckhardt, S. Klein, E. Kuehn, and W. M. Wajda, ``Energy savings in mobile networks based on adaptation to traffic statistics,'' {\em Bell Labs Tec. J.}, vol. 15, no. 2, pp. 77--94, Sep. 2010.

\bibitem{Mia10} G. Miao, N. Himayat, and G. Li, ``Energy-efficient link adaptation in frequency-selective channels,'' {\em IEEE Trans. Commun.}, vol. 58, no.2, pp. 545--554, Feb. 2010.

\bibitem{convex} S. Boyd and L. Vandenberghe, {\em Convex Optimization}, Cambridge University Press, 2004.

\bibitem{Wang11} X. Wang and G. Giannakis, ``Resource allocation for wireless multiuser OFDM networks,'' {\em IEEE Trans. Inf. Theory}, vol. 57, no. 7, pp. 4359--4372, July 2011.

\bibitem{cvx} M. Grant and S. Boyd, {\em CVX: Matlab software for discplined convex programming}, http://cvxr.com/cvx/, Oct. 2010.

\end{thebibliography}
\end{document}